\documentclass[manuscript]{rmaa}
\usepackage{paralist}
\usepackage{psfrag,color}
\usepackage[latin1]{inputenc}
\title{Quantitative Stellar Spectral Classification. IV.
       Application to the Open Cluster IC~2391}
\author{J. Garc\'{\i}a,\altaffilmark{1,2}
        N. S\'anchez,\altaffilmark{3} and
        R. Vel\'asquez\altaffilmark{4}}
\altaffiltext{1}{NASA Goddard Space Flight Center, Greenbelt, MD 20771}
\altaffiltext{2}{IACS/Department of Physics, The Catholic University of
                 America, Washington, DC 20064}
\altaffiltext{3}{Instituto de Astrof\'{\i}sica de
                 Andaluc\'{\i}a, CSIC, Spain}
\altaffiltext{4}{Centro de Modelado Cient\'{\i}fico,
                 Universidad del Zulia, Venezuela}
\shortauthor{Garc\'{\i}a et al.}
\shorttitle{Quantitative Stellar Spectral Classification. IV.}
\fulladdresses{
\item J. Garc\'{\i}a, NASA Goddard Space Flight Center, Greenbelt,
      MD 20771, US (javier@milkyway.gsfc.nasa.gov).
\item N. S\'anchez, Instituto de Astrof\'{\i}sica de Andaluc\'{\i}a,
      CSIC, Apdo. 3004, E-18080, Granada, Spain (nestor@iaa.es).
\item R. Vel\'asquez, Centro de Modelado Cient\'{\i}fico,
      Universidad del Zulia, Maracaibo, 4004, Venezuela
      (raulv@cida.ve).}
\listofauthors{J. Garc\'{\i}a, 
               N. S\'anchez, \& R. Vel\'asquez.}
\indexauthor{Garc\'{\i}a, J.}
\indexauthor{S\'anchez, N.}
\indexauthor{Vel\'asquez, R.}
\date{jue sep 18 09:39:41 EDT 2008}

\resumen{En este trabajo realizamos la primera prueba de un m\'etodo de 
clasificaci\'on espectral estelar (Stock \& Stock 1999), aplic\'andolo 
a estrellas tempranas. La muestra contiene estrellas miembros del
c\'umulo abierto IC~2391, para las cuales se dispone de espectros de
alta resoluci\'on del Projecto UVES del Observatorio Paranal.
Mostramos que, en general, las magnitudes absolutas $M_V$ y los 
colores intr\'insecos $(B-V)_0$ pueden ser recuperados dentro de los
errores predichos por la calibraci\'on original ($\sim 0,4$ para las
magnitudes y $\sim 0,03$ para los colores). Este tipo de precisi\'on nos 
permite estimar las distancias de estrellas individuales e
inferir su membres\'ia, para as\'i obtener una distancia promedio
al c\'umulo de $156 \pm 24$ pc, en muy buena concordancia con
valores reportados previamente. Finalmente discutimos las 
ventajas y dificultades de usar este 
m\'etodo y c\'omo puede ser mejorado para su implementaci\'on en
estudios futuros.}

\abstract{In this work we perform the first test of a stellar spectral
classification method (Stock \& Stock 1999) by applying it
to early type stars. The sample of stars are the members of
the open cluster IC~2391 that have high-resolution spectra
available in the UVES Project of Paranal Observatory. We
show that, in general, absolute magnitudes $M_V$ and intrinsic
colors $(B-V)_0$ can be recovered within the uncertainties
stated in the original calibration ($\sim 0.4$ for the
magnitudes and $\sim 0.03$ for the colors). This accuracy
allows us to estimate distances and to infer membership
of individual stars to obtain an average distance to the
cluster of $156\pm 24$ pc, which is in good agreement
with previous reported determinations. Finally, we identify
and discuss the real strengths and limitations of this
method and we suggest how it can be improved for future
studies.}

\addkeyword{Methods: data analysis}
\addkeyword{Stars: fundamental parameters, classification}
\addkeyword{Open cluster: individual (IC~2391)}

\begin{document}

\maketitle

\section{Introduction}

The absolute magnitude and the intrinsic color are
the main quantities to describe stars. The first
provides the amount of energy emitted, and combined
with the apparent magnitude gives an estimate of
the distance to the object. The second is directly
related to the temperature of the atmosphere of
the star. These properties are expected to imprint
themselves on the optical band of the spectrum,
specially trough the absorption lines of different
species. The spectral classification methods
are advocated to infer these parameters by the
inspection of those absorption profiles, and it is
often made by visual inspection, as is the case for 
the MK system. New {\it quantitative} classification
methods are needed to improve the results and
provide consistency in the process.
Many authors have pointed out the need of a
fully automated system to classify stellar spectra \citep[see for
example the reviews by][]{Bai02,All04,Gir06}. Automated systems are
more homogeneous (and therefore more appropriate) for the classification
of a large number of stars on the basis of a self-consistent system.

Due to the progress in astronomical instrumentation and to the
increase in the data acquisition rate, new automated and efficient
systems are constantly being developed and improved to classify
stellar spectra and/or to infer stellar atmospheric parameters
from spectral profiles \citep[recent examples
include][]{Chr02,All03,Bai05,Zha05,Rec06,Ref07}.
An example of this kind of systems is the method proposed by
\citet[][hereafter SS99]{SS99}. This method is based on the
determination of a spectral index defined by two external bands
placed at each extreme of a central band \citep{Wor94}. This
index is related with the stellar parameters through second
order polynomials. The calibration of this (and any) method
is performed by comparing the defined indices with physical
parameters of interest, such as the absolute magnitude $M_V$
or the color index $(B-V)_0$. The errors in these parameters
will propagate in the resulting calibration, so that in the
best case one would expect the results to be as accurate as
the data used for calibration.
Using the SS99 method is possible, in theory, to derive absolute
magnitudes and intrinsic colors with average uncertainties of
$\sim 0.3$ and $\sim 0.02$ magnitudes, respectively. This is
a reasonably good accuracy taking into account the simplicity
of the method, which does not require large computational
resources. Another advantage of this method is its relatively low
sensitivity to the spectral type. In general, the morphological
differences in the spectra of early-type and late-type stars
can lead to problems when defining robust indices. The results
found in SS99 showed that, on average, the residuals did not
vary significantly among stars of different spectral types
(in the range A-K). This is because the method does not need
to determine the underlying true continuum. The extension of
the method to B-type stars was done by \citet[][hereafter
SS02]{SS02}. In a subsequent paper, \citet{Mol04} applied
the same technique to a sample of stars within the spectral types
K-M. The effects of variations in the spectral resolution on
the sensitivity of the method was analyzed by \citet{Gar05}.
They showed that the method can be applied to low resolution
spectra without loss of accuracy and that small variations
in the spectral resolution would not affect the determination
of the physical parameters for early-type stars. However, their
results also established that for late-type stars the method
need to be improved.

To date the proposed method (SS99) has not been used to derive
physical parameters of any sample of stars, except for the same
stars used for the calibration. In this paper we perform the
first test of the method by its application to the stars in an
open cluster. Our main goal is to evaluate the method in order
to identify strengths and limitations. The open cluster chosen
for this study is IC~2391, since high resolution spectra for
$\sim 50$ possible members are available from the UVES Paranal
Observatory Project \citep{Bag03}. Another reason to choose
this cluster is the large amount of studies published in the
last years, which will allow us to verify the results obtained
using the SS99 method and to compare them with previous works.

This paper is organized in the following way. In
Section~\ref{metodo} we briefly explain the method
used to determine the stellar parameters. In
Section~\ref{muestra} we describe the sample of
stars over which the method is applied. The results
obtained are shown in Section~\ref{resultados}, where
a detailed analysis lead us to propose some improvements
to the method which allow us to determine the stellar
parameters and to compare them with literature values.
Finally, the main conclusions are summarized in
Section~\ref{conclusiones}.

\section{Method for obtaining stellar parameters}
\label{metodo}

It is worthwhile to start by briefly explaining the SS99
method. The first part of the method consists in to establish a series of
absorption lines that might provide information on the
stellar physical properties. For each line a spectral
index ($w$) is calculated following a procedure similar
to the one used by \citet{Wor94}. For a given line, an
inner region (the line itself) and two outer regions (one
on each side of the line) have been selected based on visual
inspection of representative spectra of each spectral type.
The complete list of lines and their limits can be found in
SS99 and SS02. A linear interpolation between the average
fluxes of the outer regions provides a local continuum (or
pseudo-continuum). The value of the index $w$ for each
absorption line is given by the integration of the inner
region:
\begin{equation}
w = \int_{\lambda_{1}}^{\lambda_{2}}
    \frac{F_{c}-F_{\lambda}}{F_{c}}\,d\lambda \ \ ,
\end{equation}
where $F_{c}$ represents the flux in the local continuum and
$F_{\lambda}$ is the flux of the line between the integration
limits.

In the previous papers (SS99, SS02) the authors used first,
second and third order polynomials to relate the defined
indices with the physical parameters to be recovered. The
calibration (i.e., the determination of the coefficients
of the polynomials) was made by the minimum square method.
The best results were obtained with second order polynomials
with three different lines as independent variables. Thus,
if $w_{1}$, $w_{2}$ and $w_{3}$ are the calculated indices
of three given lines then the polynomial relating a physical
parameter $PP$ (namely the absolute magnitude or the intrinsic
color) with these indices is given by
\begin{equation}
\label{polinomio}
\begin{array}{cl}
PP = & a_{1}+a_{2}w_1+a_{3}w_2+a_{4}w_3+a_{5}w_1^2+a_{6}w_2^2 \\
     & +a_{7}w_3^2+a_{8}w_1w_2+a_{9}w_1w_3+a_{10}w_2w_3 \ \ .
\end{array}
\end{equation}
The information associated to each combination of lines and
their coefficients $a_{i}$ ($i=1,...,10$) can be found in
SS99 and SS02.

It is important to remark that the stars used in SS99 were
divided into four different groups (numbered from 1 to 4).
This separation is based on the profile of certain lines
representative of different ranges of color. The range of
spectral types is approximately B9-A2 for stars in the
group 1, F0-G5 for group 2, G0-K0 for group 3, and K0-M5 for
group 4. Nevertheless, stars belonging to group 1 were excluded
from calibration in SS99 due to insufficient sample size. The
calibration for this group was done independently (using a
new sample of stars) and reported in SS02. Therefore, the SS99
polynomials should be used for stars belonging to groups 2, 3
and 4; while for those stars in group 1 the polynomials in
SS02 are more appropriate.

\section{The sample of stars}
\label{muestra}

The stellar spectra used in this work were obtained from
the public database of the UVES Paranal Observatory
Project\footnote{http://www.sc.eso.org/santiago/uvespop}
\citep{Bag03}. These spectra are characterized by a high
signal to noise ratio ($S/N \sim 300 - 500$ for the $V$ band),
high spectral resolution ($R=\lambda/\Delta\lambda \sim 80000$),
and a wide range of wavelengths ($3040-10400$ \AA).
The UVES library contains a total of 48 (presumed) members
of the open cluster IC~2391 with available data in the
spectral range of interest for this work. According to
the SS99 criteria, 20 of these stars were classified
into group 1, 15 into group 2, and the rest into groups
3 and 4. The later two groups were excluded from the
present analysis because these groups are more sensitive
to resolution variations and their analysis would require
a new calibration of the method using similar resolution
spectra \citep[as it was shown in][]{Gar05}. Therefore,
our preliminary sample consists of 35 stars which are
shown in Table~\ref{hrtab}. This table contains the
star identification, the Id number assigned by
\citet{Per69} (in this paper we use this number
to identify the stars), the spectral type, and the
assigned group according to SS99 criteria.

\section{Results and discussion}
\label{resultados}

\subsection{Application to the Open Cluster IC~2391}

Following the procedure described in Section~\ref{metodo},
we have calculated the indices for each one of the 19 lines
defined in SS99 (for the stars belonging to group 2), and for
each one of the 12 lines of SS02 (group 1). Once calculated
all these indices, we determined the physical parameters $M_V$
and $(B-V)_0$ using equation~\ref{polinomio} with the
coefficients given in SS99 and SS02. The obtained
results are shown all together in Tables~\ref{mvGrupo1},
\ref{mvGrupo2}, \ref{bvGrupo1} and \ref{bvGrupo2}. In
these tables we include the values of $M_V$ and $(B-V)_0$
resulting for each star when using each of the line
combinations (indicated with Roman numbers), defined in
SS02 for group 1 and in SS99 for group 2. The tables also
show, in the last two columns, the averages of all the
combinations for each star ($\overline{M_V}$ and
$\overline{(B-V)_0}$) and their respective standard
deviations ($\sigma_{M}$ and $\sigma_{0}$).

In principle, one would expect that for each star the results do
not differ significantly among different combinations. This is
because these are the best combinations of the method in the
sense that these are the combinations with the smallest
average residuals $r_{av}$ obtained from the polynomial
calibration (SS99, SS02). Nevertheless, the first thing
we can notice from Tables~\ref{mvGrupo1}-\ref{bvGrupo2} is
that some combinations of lines differ considerably from
the other ones and/or from the corresponding average value.
As a consequence, there  are relatively large standard
deviations for some of the obtained parameters. It is
remarkable, in particular, the anomalous behavior of
the combination {\sc i} in Table~\ref{mvGrupo1} which
corresponds to the combination of lines H$\delta$,
H$\gamma$, and H$\beta$ (SS02).
The value obtained for each star using combination {\sc i}
differs significantly from the values of other combinations.
This behavior is observed for all the stars in
Table~\ref{mvGrupo1}, but it is especially evident
in stars of spectral type A, for which $M_V>12$
in all the cases. These particular results are extremely
anomalous since the typical values for the absolute
magnitudes in type A stars lie between 0.3 and 2.4 
magnitudes. The values of $M_V$
listed in Table~\ref{mvGrupo2} are distributed more
uniformly among the different combinations, which
is reflected in the relatively small standard
deviations ($\sigma_M$) in comparison with the
values in Table~\ref{mvGrupo1}. However, we must
emphasize that the five higher values of $\sigma_M$
in Table~\ref{mvGrupo2} correspond to stars of
spectral type A. Regarding the values obtained for
the colors (Tables~\ref{bvGrupo1} and \ref{bvGrupo2}),
once again the most dispersed results correspond to
type A stars, i.e. these stars always have the highest
standard deviations $\sigma_0$.

From the previous analysis two conclusions can be drawn. First,
the combination {\sc i} for the absolute magnitude in SS02 exhibits
an anomalous behavior: their results differ systematically ($M_V$
always showing higher values) and significantly (in comparison
with the $\sigma_M$ values) from the results given
by the other combinations. We do not know the reason for this
problem, but the systematic nature of this behavior makes us
believe that one or more of the coefficients of the corresponding
polynomial could be wrong. To confirm this point it would be
necessary to repeat the calibration for this combination of lines, but
aside from the inherent difficulties this issue is beyond the
scope of this paper. Therefore, we have decided to exclude this
combination from the final results.\footnote{It has to be
mentioned that we have also performed all calculations including
the combination {\sc i} of the Table~\ref{mvGrupo1}. Although
the quality of the results decreases, the main conclusions
remain unchanged.} The second conclusion derived from
Tables~\ref{mvGrupo1}-\ref{bvGrupo2} is that the application
of the method to type A stars would require some kind of special
treatment. We will address this point in the following section.

\subsection{Comparison with the parameters derived from the MK-system}

In order to better understand the behavior of the method when
applied to the open cluster IC~2391, it is helpful to compare
the obtained parameters (Tables~\ref{mvGrupo1}-\ref{bvGrupo2})
with the values derived from the spectral type. For this we
use the Landolt-Bornstein tables \citep{Voi65}. We may assume
that the values derived from the MK types are the {\it expected}
values for $M_V$ and $(B-V)_0$, but we have to keep in mind
that these values are not necessarily the ``correct" ones.
In fact, the uncertainties associated with the values derived
from the MK types (based on the visual comparison of stellar
spectra), are in general larger than the uncertainties associated
with the method used in this work. Thus, although the comparison
done in this section does not provide information about
the accuracy and precision of the method, this analysis helps
to understand the behavior of the method and to identify
potential problems, as discussed below.

\begin{figure}\centering
\includegraphics[width=\columnwidth]{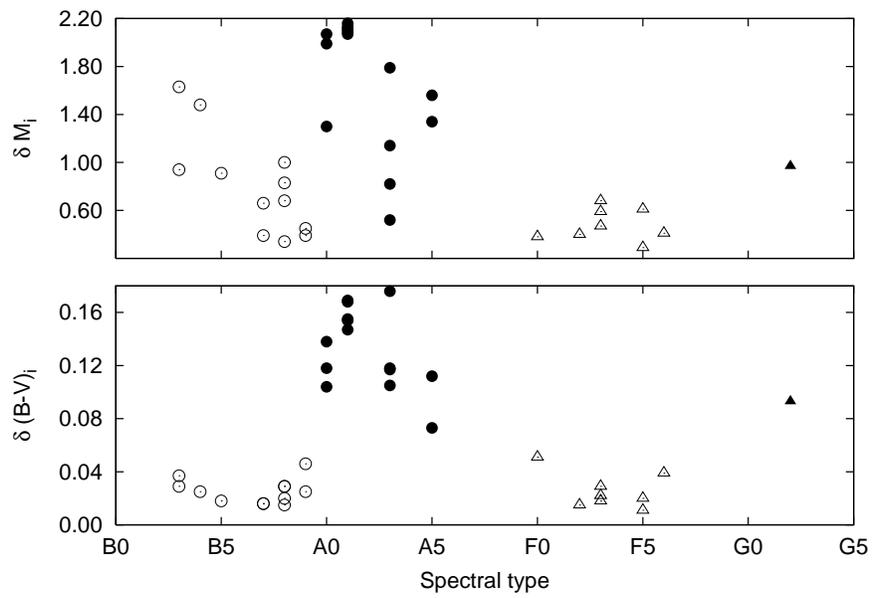}
\caption{Deviation parameter (see equation~\ref{parameter})
for the absolute magnitude (upper panel) and the intrinsic
color (lower panel) for the used stars as a function of the spectral
type. The open circles correspond to B stars, solid circles to type A,
open triangles to type F, and solid triangles to type G.}
\label{fig1}
\end{figure}

We have first calculated the absolute values of the differences
between the values obtained when applying the method
(Tables~\ref{mvGrupo1}, \ref{mvGrupo2}, \ref{bvGrupo1}
and \ref{bvGrupo2}) and the values derived from the MK types.
This is done for each star and each line combination (except
the combination {\sc i} in Table~\ref{mvGrupo1}). Then, for
each star, we determined the average value of all these
differences. In other words, if $M_V(i, j)$ denotes the
magnitude of the $i$-th star calculated by using the
$j$-th combination and if $M_V^{mk}(i)$ denotes the
magnitude derived from its spectral type, then we can
define a {\it deviation parameter} for the magnitude
in the form
\begin{equation}
\label{parameter}
\delta M_i = \frac{1}{N_c} \sum_{j=\rm{I}}^{N_c}
\left| M_V(i,j)-M_V^{mk}(i) \right| \ \ ,
\end{equation}
$N_c$ being the number of available combinations. The
deviation parameter for the color $\delta(B-V)_i$ is
calculated in a similar way. These parameters can be
zero only if all the combinations yield exactly the
same result and this result agrees with the one
derived from the MK types. Figure~\ref{fig1} shows
these deviation parameters for all the stars in the
sample as a function of the spectral type. Clearly,
most of type A stars (solid circles in Figure~\ref{fig1})
exhibit the largest deviations.
This behavior can be easily understood by considering the
small number of stars belonging to this spectral type used
in the original calibration of the method. The sample in
SS99 included 487 stars, but only 10 of them were of spectral
type A. In fact, the group 1 (that includes early type stars)
was excluded from that study because of this reason. In order
to extend the calibration to early type stars, an observing
program and calibration was carried out later (SS02). However,
that work was restricted to spectral types O-B, again
excluding the A types. Therefore, it is suitable to apply
this calibration to stars belonging to spectral types from O 
to G, but the total number of A stars
used for the final calibration (SS99 and SS02) is rather small
and thus the reliability of the results for this spectral type
is questionable.
In order to solve this problem, it will be necessary to
recalibrate the method using a sample of stars that covers
a wider and continuous range of spectral types.
In this work we have calculated the physical parameters
for all the available stars but we have rejected type
A stars when estimating the final set of parameters for
the open cluster IC~2391.

Now we proceed to calculate a deviation parameter for each
combination of lines instead of for each star. To do this,
we take the average of all the available stars in a similar
way to equation~\ref{parameter}. For the magnitude we define
\begin{equation}
\label{noabs}
\delta M_j = \frac{1}{N_s} \sum_{i=1}^{N_s}
M_V(i,j)-M_V^{mk}(i) \ \ ,
\end{equation}
where $N_s$ represents the number of available stars (and
similar for the intrinsic color). Note that in this equation
we are not taking absolute value of the difference. We perform the
calculations for all the combinations, including
the combination {\sc i} in Table~\ref{mvGrupo1},
but now we exclude type A stars due to the reasons
mentioned previously. The results are shown in
Figure~\ref{fig2}. The error bars in this figure
represent the corresponding standard deviations.
These error bars include both the uncertainties of
the method itself and the uncertainties in the
parameters derived from the MK system. Therefore,
these error bars do not indicate the degree of
uncertainty associated with a given combination,
but they are a meaningful measure of ``quality"
of a combination relative to others. We see in
Figure~\ref{fig2}a that, as mentioned before,
the worst combination for the magnitude is
the combination {\sc i}. This combination exhibits
the largest average difference ($\sim 3$ magnitudes),
and its standard deviation is significantly larger
than the others. The rest of the combinations, both
for $M_V$ and for $(B-V)_0$, show a similar
behavior: the average differences are distributed
around zero (within the uncertainties), and their
standard deviations are nearly the same.

\subsection{A strategy to improve the determination of the stellar 
parameters}

\begin{figure}[!t]
\includegraphics[scale=0.9]{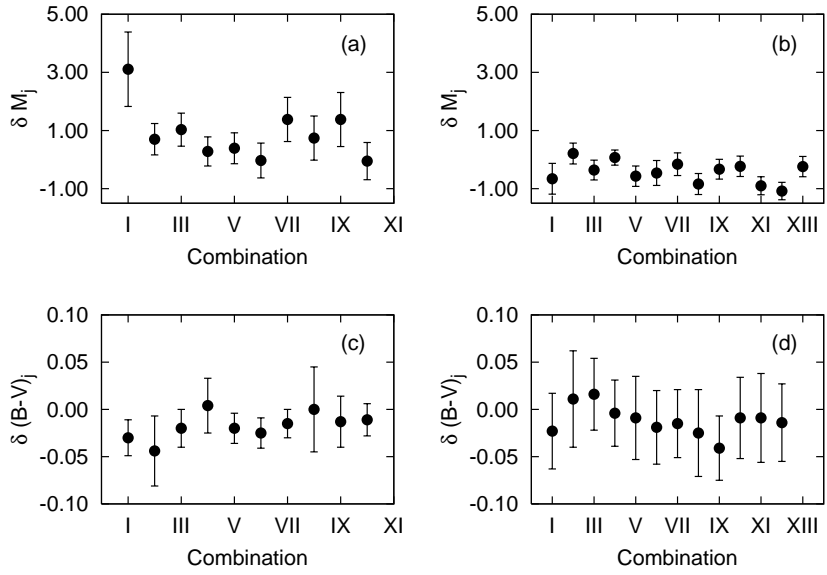}
\caption{Deviation parameter (see equation~\ref{noabs}) calculated 
for each combination of lines defined in SS99 and SS02.
Panel (a) refers to the absolute magnitudes of group 1 stars,
(b) to the absolute magnitudes of stars in group 2, (c) to
intrinsic colors of group 1, and (d) to the colors of group 2.
The bars indicate the standard deviation in each case.}
\label{fig2}
\end{figure}

In Figure~\ref{fig2} we can see that some combinations
(apart from combination {\sc i} in Figure~\ref{fig2}a)
exhibit standard deviations larger than those found for
the rest. As an example, we can mention the combinations
{\sc ii} and {\sc viii} in Figure~\ref{fig2}c. In
principle, one would expect to find less accurate
(even biased) results when these combinations are
used. These results do not necessarily agree with
the SS99 and SS02 results. For example, combination
{\sc ii} in Figure~\ref{fig2}c is the ``best" one
(i.e., that having the smallest average residual
$r_{av}$) according to SS02. In any case, the
residual coming from the calibration with a
sample of stars is not the same thing as the
expected uncertainty when applying a given
combination to a star. What combination(s)
would we expect to provide the best results?
This is not a trivial question. It seems from Figure~\ref{fig2}
that most of the combinations behave quite similarly. However,
the choice of a particular combination may have undesirable
consequences on the final results if this combination works
worse for the particular sample of stars that is used. This
is the case, most likely by chance, of combination {\sc ii}
in Figure~\ref{fig2}c.

\begin{figure}[!t]
\includegraphics[scale=0.9]{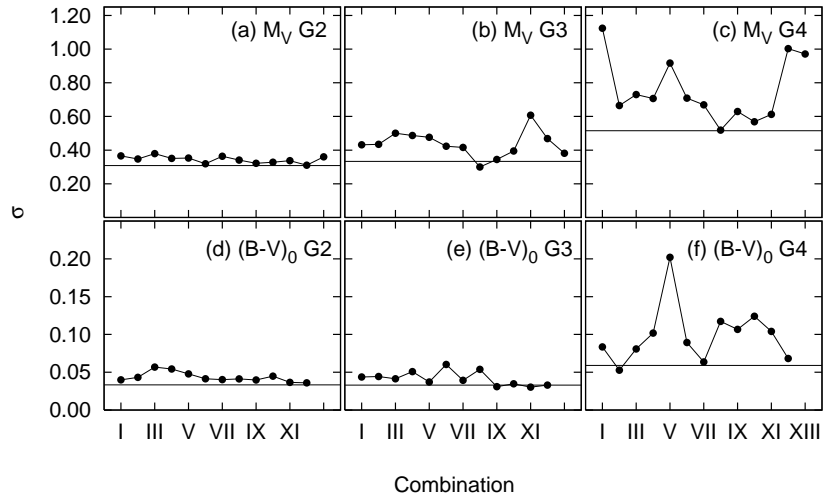}
\caption{Standard deviations of the differences between the values
assigned to the parameters recovered from the polynomials for the
stars used in SS99. Each panel refers to the result of a given 
parameter ($M_V$ o $(B-V)_0$), and to each of the groups (2, 3 and 4)
used in SS99. The horizontal lines indicate the values
obtain by using the averages of all the combinations instead of a single
one.}
\label{fig3}
\end{figure}

To solve this problem we propose to average the results
obtained with the different combinations. This simple step
will diminish any systematic deviation caused by the
application of a given combination to a (small) sample
of stars such as IC~2391. In order to properly evaluate
the suitability of this proposal, we have calculated
absolute magnitudes and intrinsic colors for the same
sample of stars used in SS99. Afterwards, we have compared
the magnitudes and colors used for the calibration with the
values obtained from the polynomials (we have used polynomials
with three lines as independent variables to be consistent with
the results presented here).
Figure~\ref{fig3} shows the standard deviations ($\sigma$)
resulting from the differences between these two values for
each combination and for stars belonging to the groups 2, 3,
and 4. The solid horizontal lines in these figures indicate
the values obtained when using the average of all the 
combinations. The values shown are larger than those
reported in SS99 because in this case we are using all the stars
in the sample and not only the ones used for the calibration
of the polynomials. In fact, these are the standard deviations
that determine the uncertainties associated with each polynomial.
For the absolute magnitudes the standard deviations are around
$\sigma \sim 0.4$ for groups 2 and 3 but, as expected, are higher
(in the range $0.6 \lesssim \sigma \lesssim 1.0$) for group 4.
The same holds for the intrinsic colors, i.e., the method
behaves worse for late type stars belonging to group 4. The
important point here is that when the averages of the combinations
are used (horizontal lines in Figure~\ref{fig3}), the stellar
parameters can be recovered with uncertainties of the order of
those associated with the best combinations. If
we select the combination {\sc v} to determine the color of
a star belonging to group 4, then we obtain on average, an
uncertainty $\sim 4$ times higher than if we use the average
of all the combinations.

Obviously, this effect has a minor impact on early type stars, but this
strategy (to use the averages of the results derived from all the
available combinations instead of from only one combination) is
highly recommended. By choosing a single combination three spectral
indices are being used. If at least one of these indices is not
well determined because there is a peculiar feature in the
spectrum (e.g., a near emission line), then the value obtained
from the polynomial could be significantly affected. In general,
the selection of the polynomial is based on the presence or absence
of certain lines in the observed spectrum. What we are suggesting is
to select all the available combinations and to average the
results because this procedure minimizes the effects of random
errors. This is the procedure which we will use to calculate
the final values for the stars in our sample, that will be 
discussed in the next section.

\subsection{Final parameters for the stars in IC~2391}

According to the previously explained arguments, we have excluded
from the analysis both the type A stars and the combination {\sc i}
of Table~\ref{mvGrupo1}. Next we have determined the absolute
magnitudes $M_V$ and the intrinsic colors $(B-V)_0$ for the
sample of stars by averaging the values obtained for each
combination. These results are listed in Table~\ref{parfinales}.
The columns contain the following information: the identification
(Id) of each star according to \citet{Per69}, the values obtained
for each parameter ($M_V$ and $(B-V)_0$), the values derived
from the MK types (indicated with the superscript $mk$), and
the values from different sources in the literature (indicated
with the superscript $lit$).
Note that $M_V^{mk}$ and $M_V^{lit}$ can differ significantly.
We have calculated the absolute values of the differences between
the values from the MK system and from the literature and those
calculated in this work. The averages of these differences
are $0.59$ and $0.35$ magnitudes when we compare our results
with $M_V^{mk}$ and $M_V^{lit}$, respectively. Our results are
more consistent with the ones reported in the literature
indicated in Table~\ref{parfinales}. The $M_V^{mk}$ values
are homogeneous in the sense that they come from only one
source (the MK system), and this is the reason why we have
used MK to analyze our results. Nevertheless, we have already
pointed out that uncertainties associated with $M_V^{mk}$
can be quite large, and this is probably the main reason
for the differences with our results. Regarding the colors,
the average of the differences between our results and
$(B-V)_0^{mk}$ or $(B-V)_0^{lit}$ are $0.023$ or $0.029$
magnitudes, respectively. Note that the average residual 
$r_{av}$ for $M_V$ is
around $0.30$ magnitudes in SS99 and $0.40$ magnitudes in
SS02, whereas for $(B-V)_0$ is of the order of $0.02$
magnitudes in both works. The average differences obtained
are in good agreement with these values.

\subsection{The distance to the open cluster IC~2391}

Using the apparent magnitudes given in the web page of the UVES
Paranal Observatory Project and neglecting interstellar absorption,
we have estimated the distance modulus and hence the distance for
each star in the sample. Table~\ref{distancia} shows the derived
distance values ($d_{\rm{mod}}$), including the errors associated 
with the method, i.e. the uncertainties propagated from
the average residuals of the polynomial calibration. It also shows, 
for comparison, the distances and their errors derived from parallaxes
given in the Hipparcos
catalog ($d_{\rm{par}}$), which were obtained from the SIMBAD
database. The last column shows the percentage difference
($100\times|d_{\rm{mod}} - d_{\rm{par}}|/d_{\rm{par}}$) for
those stars having both distances available. The majority of
stars have small differences between $d_{\rm{mod}}$ and
$d_{\rm{par}}$ ($\lesssim 10\%$), although there are three
stars with relatively large differences ($\gtrsim 30\%$).

In order to obtain the best estimation of the distance to the open
cluster IC~2391, we have used the following procedure. First, we calculate
the distance of each star. Then we compute the average value of the
distances and the standard deviation. Stars having distances deviating
more than $2.5$ times the standard deviation from the average are
rejected, and the procedure is repeated until no further stars
can be rejected. At the end of this procedure, five stars were
discarded from distance estimation: 1, 17, 36, 45, and 46. The
value of the distance obtained for the open cluster IC~2391 and its
standard deviation is $d = 156 \pm 24\ \mathrm{pc}$. The three
stars having the largest differences with this value are the
stars numbered 36, 45, and 46. After reviewing the available
literature, we note that several authors have suggested
non-membership of these stars to IC~2391
\citep{Per69, PerBon69, Mai86}.
The distances of stars 1 and 17 are outside $2.5$ times the
standard deviation but, furthermore, their values are in
good agreement with the ones derived from the parallaxes
(Table~\ref{distancia}). \citet{Bus65} identified star 1
as a member of IC~2391, whereas \citet{Lev88} and \citet{Dod04}
classified it as a non-member. For the case of the star 17,
both \citet{Per69} and \citet{PerBon69} classified it as
a non-member of the cluster. In a recent work, \citet{Pla07}
concluded that none of the five rejected stars are member of
IC~2391.

The distance to IC~2391 obtained in this work ($d = 156\ \mathrm{pc}$)
is in very good agreement with the values given by other authors.
\citet{Bar01} estimated a distance of $155$ pc, \citet{Dod04}
derived $147.5$ pc, and \citet{Pla07} obtained $159.2$ pc.
This good agreement allows us to conclude that, in general,
the method used in this work (SS99, SS02) can be applied with
confidence to stellar spectra to derive their properties within
acceptable uncertainties.

\section{Conclusions}
\label{conclusiones}

In this work we have applied the method proposed by SS99 and SS02
to a sample of stellar spectra available for the open cluster
IC~2391. We have calculated the absolute magnitudes $M_V$
and intrinsic colors $(B-V)_0$ for all the studied stars
(Tables~\ref{mvGrupo1}-\ref{bvGrupo2}), as well as
the distance to the cluster. Both the detailed
analysis performed in this work and the comparison of our
results with other results from the literature allowed us
to evaluate more comprehensively the strengths and limitations
(and therefore the potential applicability) of this method.

From our results we can draw the following general conclusions:

\begin{itemize}
\item In the method described by SS99 and SS02, the combinations
of lines having the smallest average residuals are not necessarily
the ones yielding the best results for $M_V$ and $(B-V)_{0}$. In
fact, the average uncertainties may vary significantly among the
different combinations, mainly in late type stars. In this work
we propose to use the averages of the results derived from all
the available combinations instead of from only one combination.
We have verified that this procedure reduces considerably the
overall uncertainty.

\item The method used does not provide reliable results (it
generates relatively large uncertainties) when applied to
type A stars. The problem seems to lie in the fact that
this spectral type was not adequately represented in the
original calibration (SS99, SS02). It becomes necessary,
therefore, a new calibration of the method using a more
extensive spectral library.

\item From the results obtained for this sample of stars, we
conclude that it is possible to derive absolute magnitudes with
uncertainties of $0.59$ or $0.35$ magnitudes, depending on whether
the comparison is done with values derived from the spectral type
or values reported in the literature, respectively. It is also
possible to derive intrinsic colors with average uncertainties
of $0.023$ or $0.029$ magnitudes, depending again on the source
used for the comparison. These uncertainties are in agreement
with the average residuals associated to $M_V$ and $(B-V)_{0}$
according to SS99 and SS02.

\item Stars numbered 1, 17, 36, 45 and 46 are probably not
members of IC~2391. The rest of the stars allowed us to
estimate a distance of $156 \pm 24\ \mathrm{pc}$ for the
cluster, which is in good agreement with published values.
\end{itemize}

Summarizing, this work shows that the method proposed in SS99
and SS02 can be applied in a reliable way to calculate the
physical parameters of stellar systems. Both the simplicity
in the definition of the indices used and the relatively
small uncertainties in magnitude and color values make
this method a suitable tool for the analysis of a large
number of stellar spectra. Nevertheless, a new calibration
is necessary in order to overcome the current limitations
of the method, mainly those associated with type A stars.

\acknowledgments{NS acknowledges financial support from MEC of Spain through grant
AYA2007-64052 and from Consejer\'{\i}a de Educaci\'on y Ciencia
(Junta de Andaluc\'{\i}a) through TIC-101.}



\begin{table}[!t]\centering
\caption{Stars in the open cluster IC~2391 used in 
this work according to their corresponding group (SS99).}
\label{hrtab}
\begin{tabular}{lcl @{\hspace{2cm}} lcl}
\toprule
\multicolumn{6}{c}{Group 1}\\
Star     & Id & Spec. Type & Star  & Id & Spec. Type \\
\midrule
HD 73287   &  1 & B7V      & HD 74275   & 23 & A0V        \\
HD 73503   &  2 & A0V      & HD 74516   & 29 & A1V        \\
HD 73681   &  3 & A1V      & HD 74535   & 31 & B8s        \\
HD 73952   &  8 & B8Vn     & HD 74560   & 34 & B3IV       \\
HD 74146   & 16 & B4IV     & HD 74955   & 41 & A1V        \\
HD 74168   & 17 & B9p      & HD 74999   & 42 & A1V        \\
HD 74071   & 13 & B5V      & HD 75067   & 45 & B8/B9IV/Vn \\
HD 74196   & 21 & B7Vn     & HD 75105   & 46 & B8III/IV   \\
HD 74195   & 20 & B3IV     & HD 75184   & 47 & A1V        \\
HD 74196   & 21 & B7Vn     & HD 75466   & 50 & B8V        \\
\midrule
\multicolumn{6}{c}{Group 2}\\
Star   & Id & Spec. Type & Star  & Id & Spec. Type \\
\midrule
HD 73722   &  4 & F5V      & HD 74582   & 36 & F3III      \\
HD 73778   &  5 & F0V      & HD 74762   & 40 & A5V        \\
HD 74009   & 10 & F3V      & HD 75029   & 43 & A2/A3m     \\
HD 74044   & 11 & A3m      & HD 75202   & 49 & A3IV       \\
HD 74117   & 14 & F2IV/V   & IC 2391-H5 & 51 & F5V        \\
HD 74182   & 19 & A5IV     & IC 2391-52 & 52 & F6V        \\
HD 74537   & 33 & A3IV     & SHJM 2     & 73 & G2         \\
HD 74561   & 35 & F3V      \\
\bottomrule
\end{tabular}
\end{table}


\begin{table}[!t]\centering
\setlength{\tabcolsep}{4.5pt}
\begin{changemargin}{-2cm}{-2cm}
\caption{Absolute magnitudes for the stars in group 1 obtained
with each combination defined in SS02.}
\label{mvGrupo1}
\begin{tabular}{*{13}{c}}
\toprule
Id &\sc{i}&\sc{ii}&\sc{iii}&\sc{iv}&\sc{v}&\sc{vi}&\sc{vii}&\sc{viii}&
\sc{ix}&\sc{x} &$\overline{M_V}$ & $\sigma_{M}$ \\ 
\midrule
 1  & 2.52 & 0.13  & 0.70  &-0.36 &-0.17 &-0.77 & 
0.64 & 0.03 & 0.43 &-0.74  & 0.24 & 0.95 \\
 2  &12.17 & 3.88  & 4.77  & 1.94 & 3.09 & 0.16 & 
2.65 &-0.82 & 3.61 & 0.04  & 3.15 & 3.66 \\
 3  &12.66 & 3.88  & 4.87  & 0.95 & 3.09 & 0.28 & 
3.19 & 0.00 & 6.40 & 0.37  & 3.57 & 3.85 \\
 8  & 6.06 & 1.08  & 1.68  & 0.39 & 0.62 &-0.34 & 
1.37 & 0.13 & 1.70 &-0.43  & 1.23 & 1.86 \\
13  & 1.70 &-0.43  &-0.07  &-0.86 &-0.76 &-1.13 & 
0.55 & 0.16 & 1.05 &-1.09  &-0.09 & 0.96 \\
16  & 1.60 &-0.62  &-0.55  &-0.98 &-0.84 &-1.16 & 
0.34 &-0.34 & 0.07 &-1.03  &-0.35 & 0.84 \\
17  & 3.32 & 0.34  & 0.71  &-0.24 &-0.01 &-0.51 & 
0.79 & 0.08 & 0.74 &-0.59  & 0.46 & 1.12 \\
18  & 6.60 & 1.88  & 2.64  & 1.11 & 1.42 & 0.08 & 
2.31 & 0.29 & 3.77 & 0.20  & 2.03 & 1.99 \\
20  &-0.40 &-1.74  &-1.86  &-2.04 &-2.00 &-1.85 &
-0.35&-1.07 &-0.35 &-1.92  &-1.36 & 0.74 \\
21  & 1.34 &-0.48  &-0.02  &-0.54 &-0.74 &-1.10 & 
0.15 &-0.17 &-0.11 &-1.08  &-0.28 & 0.71 \\
23  &12.41 & 3.93  & 4.84  & 0.98 & 3.11 & 0.16 & 
2.61 &-0.78 & 3.71 &-0.02  & 3.10 & 3.78 \\
29  &13.50 & 4.15  & 5.12  & 2.16 & 3.33 & 0.16 & 
2.79 &-1.00 & 4.03 &-0.02  & 3.42 & 4.07 \\
31  & 1.35 &-0.34  &-0.12  &-0.81 &-0.59 &-0.71 & 
0.22 &-0.09 & 0.02 &-0.79  &-0.19 & 0.65 \\
34  & 0.29 &-1.08  &-1.15  &-1.44 &-1.40 &-1.35 & 
0.29 &-0.23 & 0.78 &-1.35  &-0.66 & 0.86 \\
42  &12.49 & 3.68  & 4.51  & 2.46 & 2.77 & 0.17 & 
2.89 & 0.14 & 6.37 & 0.21  & 3.57 & 3.73 \\
41  &12.26 & 3.92  & 4.64  & 1.94 & 3.05 & 0.25 & 
2.93 & 0.07 & 6.23 & 0.37  & 3.57 & 3.66 \\
45  & 2.74 & 0.32  & 0.73  & 0.15 &-0.03 &-0.63 & 
0.57 & 0.03 & 0.47 &-0.71  & 0.36 & 0.96 \\
46  & 1.95 &-0.15  & 0.31  &-0.66 &-0.48 &-0.88 & 
0.33 &-0.07 & 0.15 &-0.94  &-0.04 & 0.84 \\
47  &13.20 & 4.04  & 4.98  & 2.07 & 3.22 & 0.30 & 
2.72 &-0.91 & 3.96 &-0.02  & 3.36 & 3.96 \\
50  & 4.31 & 0.91  & 1.54  & 0.30 & 0.53 &-0.45 & 
1.12 &-0.12 & 1.10 &-0.43  & 0.88 & 1.38 \\
\bottomrule
\end{tabular}
\end{changemargin}
\end{table}


\begin{table}[!t]\centering
\setlength{\tabcolsep}{4.5pt}
\begin{changemargin}{-2cm}{-2cm}
\caption{Absolute magnitudes for the stars in group 2 obtained
with each combination defined in SS99.}
\label{mvGrupo2}
\begin{tabular}{*{16}{c}}
\toprule
Id&\sc{i}&\sc{ii}&\sc{iii}&\sc{iv}&\sc{v}&\sc{vi}&\sc{vii}&\sc{viii}&
\sc{ix}&\sc{x}&\sc{xi}&\sc{xii}&\sc{xiii}& 
$\overline{M_V}$ & $\sigma_{M}$ \\
\midrule
 4&2.38 &3.21 &2.65 &3.27 &2.77 & 3.17 & 3.14 & 2.70 & 
3.15 & 3.04 & 2.63 & 2.44 & 3.06 & 2.89&  0.31 \\
 5&2.90 &3.29 &2.87 &3.04 &2.48 & 2.99 & 3.08 & 2.23 & 
2.95 & 2.95 & 2.08 & 1.80 & 2.96 & 2.74&  0.45 \\
10&2.79 &3.81 &3.13 &3.39 &3.13 & 2.86 & 3.13 & 2.71 & 
3.02 & 3.06 & 2.64 & 2.57 & 2.98 & 3.02&  0.33 \\
11&1.33 &0.93 &1.18 &0.79 &0.43 & 0.04 &-0.29 &-0.79 & 
1.69 & 1.24 & 1.08 & 1.22 & 0.39 & 0.71&  0.72 \\
14&2.82 &3.09 &2.88 &3.07 &2.38 & 2.65 & 3.14 & 2.11 & 
2.76 & 3.01 & 1.92 & 1.65 & 2.97 & 2.65&  0.49 \\
19&3.49 &2.19 &3.65 &2.91 &3.11 & 1.99 & 2.18 & 1.83 & 
2.96 & 3.24 & 2.65 & 2.12 & 2.06 & 2.64&  0.62 \\
33&2.90 &2.15 &3.36 &1.30 &1.38 &-0.14 &-0.24 &-0.60 & 
2.66 & 2.68 & 1.97 & 1.91 & 0.46 & 1.52&  1.30 \\
35&2.98 &3.59 &3.19 &3.56 &2.95 & 3.06 & 3.44 & 2.70 & 
3.19 & 3.37 & 2.60 & 2.44 & 3.33 & 3.11&  0.36 \\
36&1.01 &2.06 &1.18 &1.84 &0.78 & 0.71 & 1.54 & 0.73 & 
0.83 & 0.99 & 0.42 & 0.44 & 1.48 & 1.08&  0.52 \\
40&3.33 &0.83 &3.75 &3.10 &2.93 &-0.55 &-0.80 &-0.62 & 
2.96 & 2.86 & 2.61 & 2.12 & 0.38 & 1.63&  1.80 \\
43&2.30 &2.27 &2.44 &2.01 &1.12 & 1.24 & 1.47 & 0.48 & 
2.06 & 2.36 & 1.36 & 1.38 & 1.60 & 1.70&  0.59 \\
49&3.53 &0.33 &4.21 &2.46 &2.78 &-0.91 &-1.02 &-0.64 & 
3.15 & 3.27 & 2.88 & 2.62 & 0.27 & 1.76&  1.87 \\
51&3.11 &4.30 &3.61 &4.08 &3.56 & 3.39 & 3.63 & 3.14 & 
3.56 & 3.68 & 3.17 & 2.94 & 3.58 & 3.52&  0.38 \\
52&2.91 &3.46 &3.15 &3.60 &3.10 & 3.36 & 3.61 & 2.94 & 
3.41 & 3.61 & 2.91 & 2.77 & 3.53 & 3.26&  0.31 \\
73&3.06 &5.00 &4.02 &4.70 &3.57 & 3.57 & 3.75 & 3.06 & 
4.02 & 4.06 & 3.38 & 3.09 & 3.80 & 3.78&  0.60 \\
\bottomrule
\end{tabular}
\end{changemargin}
\end{table}


\begin{table}[!t]\centering
\setlength{\tabcolsep}{3.5pt}
\begin{changemargin}{-2cm}{-2cm}
\caption{Intrinsic colors for the stars in group 1 obtained
with each combination defined in SS02.}
\label{bvGrupo1}
\begin{tabular}{{c}*{11}{c}{c}}
\toprule
Id & \sc{i} & \sc{ii} & \sc{iii} & \sc{iv} & \sc{v} & \sc{vi} &
\sc{vii} & \sc{viii} & \sc{ix} & \sc{x} & 
$\overline{(B-V)}_0$ & $\sigma_{0}$ \\
\midrule
   1 &-0.167&-0.141&-0.152&-0.148&-0.149&-0.144&
-0.141&-0.143&-0.141&-0.136&-0.146 &0.009 \\
   2 &-0.254&-0.125&-0.194&-0.046&-0.096&-0.087&
-0.118&-0.031&-0.187&-0.101&-0.124 &0.069 \\
   3 &-0.268&-0.279&-0.186&-0.054&-0.146&-0.102&
-0.136&-0.045&-0.237&-0.131&-0.158 &0.083 \\
   8 &-0.173&-0.139&-0.141&-0.052&-0.118&-0.125&
-0.114&-0.039&-0.125&-0.110&-0.114 &0.040 \\
  13 &-0.188&-0.195&-0.176&-0.183&-0.196&-0.193&
-0.184&-0.199&-0.183&-0.182&-0.188 &0.007 \\
  16 &-0.203&-0.267&-0.172&-0.203&-0.174&-0.188&
-0.167&-0.188&-0.123&-0.159&-0.184 &0.037 \\
  17 &-0.130&-0.129&-0.128&-0.052&-0.118&-0.125&
-0.113&-0.039&-0.120&-0.111&-0.107 &0.033 \\
  18 &-0.194&-0.301&-0.156&-0.090&-0.146&-0.133&
-0.146&-0.107&-0.209&-0.147&-0.163 &0.060 \\
  20 &-0.208&-0.323&-0.204&-0.223&-0.233&-0.244&
-0.227&-0.276&-0.221&-0.215&-0.237 &0.036 \\
  21 &-0.165&-0.141&-0.146&-0.148&-0.144&-0.144&
-0.143&-0.143&-0.144&-0.142&-0.146 &0.007 \\
  23 &-0.248&-0.125&-0.181&-0.046&-0.146&-0.102&
-0.133&-0.031&-0.235&-0.135&-0.138 &0.071 \\
  29 &-0.277&-0.125&-0.194&-0.046&-0.130&-0.102&
-0.141&-0.031&-0.260&-0.144&-0.145 &0.081 \\
  31 &-0.114&-0.129&-0.121&-0.052&-0.118&-0.125&
-0.114&-0.039&-0.115&-0.115&-0.104 &0.031 \\
  34 &-0.208&-0.287&-0.199&-0.214&-0.233&-0.244&
-0.222&-0.260&-0.211&-0.210&-0.229 &0.028 \\
  42 &-0.235&-0.279&-0.172& 0.029&-0.115&-0.102&
-0.123& 0.108&-0.195&-0.120&-0.120 &0.116 \\
  41 &-0.234&-0.279&-0.167&-0.015&-0.165&-0.123&
-0.150& 0.026&-0.289&-0.158&-0.155 &0.102 \\
  45 &-0.126&-0.135&-0.125&-0.084&-0.108&-0.125&
-0.114&-0.084&-0.117&-0.113&-0.113 &0.017 \\
  46 &-0.140&-0.139&-0.140&-0.091&-0.147&-0.138&
-0.138&-0.084&-0.146&-0.139&-0.130 &0.023 \\
  47 &-0.265&-0.125&-0.168&-0.041&-0.132&-0.093&
-0.129&-0.031&-0.248&-0.139&-0.137 &0.076 \\
  50 &-0.152&-0.113&-0.147&-0.117&-0.121&-0.120&
-0.119&-0.126&-0.125&-0.110&-0.125 &0.014 \\
\bottomrule
\end{tabular}
\end{changemargin}
\end{table}


\begin{table}[!t]\centering
\setlength{\tabcolsep}{3.5pt}
\begin{changemargin}{-2cm}{-2cm}
\caption{Intrinsic colors for the stars in group 2 obtained
with each combination defined in SS99.}
\label{bvGrupo2}
\begin{tabular}{ccccccccccccccc}
\toprule
Id&\sc{i}&\sc{ii}&\sc{iii}&\sc{iv}&\sc{v}&\sc{vi}&\sc{vii}&
\sc{viii}&\sc{ix}&\sc{x}&\sc{xi}&\sc{xii}&
$\overline{(B-V)}_0$ & $\sigma_{0}$ \\
\midrule
   4& 0.414& 0.429& 0.437& 0.436& 0.419& 0.421& 0.416& 
0.413& 0.392& 0.425& 0.416& 0.419& 0.420& 0.012 \\
   5& 0.337& 0.362& 0.375& 0.357& 0.357& 0.351& 0.344& 
0.345& 0.317& 0.354& 0.359& 0.350& 0.351& 0.014 \\
  10& 0.332& 0.382& 0.388& 0.351& 0.349& 0.345& 0.352& 
0.346& 0.314& 0.359& 0.371& 0.347& 0.353& 0.020 \\
  11& 0.210& 0.186& 0.248& 0.227& 0.217& 0.206& 0.161& 
0.178& 0.028& 0.211& 0.188& 0.199& 0.188& 0.055 \\
  14& 0.341& 0.389& 0.369& 0.349& 0.359& 0.337& 0.344& 
0.338& 0.310& 0.358& 0.365& 0.347& 0.351& 0.020 \\
  19& 0.258& 0.294& 0.296& 0.286& 0.251& 0.286& 0.266& 
0.224& 0.202& 0.256& 0.274& 0.254& 0.262& 0.028 \\
  33& 0.228& 0.169& 0.238& 0.214& 0.193& 0.200& 0.143& 
0.123& 0.022& 0.175& 0.178& 0.219& 0.175& 0.059 \\
  35& 0.394& 0.431& 0.417& 0.406& 0.407& 0.388& 0.398& 
0.386& 0.368& 0.404& 0.405& 0.400& 0.400& 0.016 \\
  36& 0.362& 0.439& 0.421& 0.371& 0.380& 0.358& 0.361& 
0.358& 0.348& 0.381& 0.376& 0.376& 0.378& 0.027 \\
  40& 0.257& 0.227& 0.254& 0.256& 0.231& 0.230& 0.176& 
0.150& 0.051& 0.199& 0.199& 0.245& 0.206& 0.059 \\
  43& 0.235& 0.232& 0.287& 0.268& 0.265& 0.257& 0.224& 
0.219& 0.143& 0.255& 0.245& 0.258& 0.241& 0.037 \\
  49& 0.242& 0.196& 0.257& 0.227& 0.213& 0.206& 0.156& 
0.097& 0.015& 0.165& 0.183& 0.285& 0.187& 0.073 \\
  51& 0.435& 0.449& 0.475& 0.455& 0.448& 0.431& 0.437& 
0.421& 0.413& 0.439& 0.440& 0.440& 0.440& 0.016 \\
  52& 0.414& 0.444& 0.440& 0.439& 0.429& 0.429& 0.430& 
0.428& 0.401& 0.436& 0.424& 0.425& 0.428& 0.012 \\
  73& 0.531& 0.534& 0.583& 0.567& 0.533& 0.535& 0.544& 
0.502& 0.533& 0.530& 0.523& 0.534& 0.537& 0.021 \\
\bottomrule
\end{tabular}
\end{changemargin}
\end{table}


\begin{table}[!t]\centering
\caption{Absolute magnitudes and intrinsic colors
for the stars in IC~2391.}
\label{parfinales}
\begin{tabular}{crrrrrr}
\toprule
 Id & $M_{V}$ & $M_{V}^{mk} $ & $M_{V}^{lit}$
& $(B-V)_{0}$ & $(B-V)_0^{mk}$ & $(B-V)_0^{lit}$ \\
\midrule
1  & -0.01 & -0.60 & -0.40$^{\rm{d}}$ & -0.146 & -0.130 &  .... \\ 
8  &  0.69 & -0.25 &  0.30$^{\rm{a}}$ & -0.114 & -0.110 & -0.100$^{\rm{a}}$ \\
13 & -0.29 & -1.20 & -0.60$^{\rm{a}}$ & -0.188 & -0.170 & -0.150$^{\rm{a}}$ \\
16 & -0.57 & -2.05 & -0.70$^{\rm{a}}$ & -0.184 & -0.185 & -0.140$^{\rm{a}}$ \\
17 &  0.15 &  0.20 &  0.00$^{\rm{a}}$ & -0.107 & -0.070 & -0.120$^{\rm{a}}$ \\
20 & -1.46 & -2.40 & -1.80$^{\rm{a}}$ & -0.237 & -0.200 & -0.180$^{\rm{a}}$ \\
21 & -0.45 & -0.60 & -0.80$^{\rm{a}}$ & -0.146 & -0.130 & -0.140$^{\rm{a}}$ \\
31 & -0.36 & -0.25 & -0.50$^{\rm{a}}$ & -0.104 & -0.110 & -0.160$^{\rm{a}}$ \\
34 & -0.77 & -2.40 & -1.20$^{\rm{a}}$ & -0.229 & -0.200 & -0.170$^{\rm{a}}$ \\
45 &  0.10 &  0.24 & -0.30$^{\rm{a}}$ & -0.113 & -0.090 & -0.100$^{\rm{a}}$ \\
46 & -0.27 & -0.95 & -0.40$^{\rm{a}}$ & -0.130 & -0.110 & -0.140$^{\rm{a}}$ \\
50 &  0.50 & -0.25 &  0.30$^{\rm{a}}$ & -0.125 & -0.110 & -0.100$^{\rm{a}}$ \\
4  &  2.89 &  3.50 &  3.80$^{\rm{a}}$ &  0.420 &  0.440 &  0.430$^{\rm{a}}$ \\
5  &  2.74 &  2.70 &  2.70$^{\rm{b}}$ &  0.351 &  0.300 &  0.360$^{\rm{b}}$ \\
10 &  3.02 &  3.57 &  3.60$^{\rm{a}}$ &  0.353 &  0.380 &  0.410$^{\rm{a}}$ \\
14 &  2.65 &  3.00 &  3.40$^{\rm{a}}$ &  0.351 &  0.350 &  0.360$^{\rm{a}}$ \\
35 &  3.11 &  3.57 &  3.33$^{\rm{c}}$ &  0.400 &  0.380 &  .... \\ 
36 &  1.08 &  1.67 &  ....            &  0.378 &  0.377 &  .... \\ 
51 &  3.52 &  3.50 &  ....            &  0.440 &  0.440 &  .... \\ 
52 &  3.26 &  3.67 &  ....            &  0.428 &  0.467 &  .... \\ 
73 &  3.78 &  4.70 &  4.25$^{\rm{b}}$ &  0.537 &  0.630 &  0.570$^{\rm{b}}$ \\
\hline
 & \multicolumn{6}{l}{(a) Perry \& Hill(1969)} \\
 & \multicolumn{6}{l}{(b) Eggen (1991)}\\
 & \multicolumn{6}{l}{(c) Patten \& Simon (1993)} \\
 & \multicolumn{6}{l}{(d) Buscombe (1965)} \\
\bottomrule
\end{tabular}
\end{table}


\begin{table}[!t]\centering
\caption{Estimated distances for the stars in IC~2391.}
\label{distancia}
\begin{tabular}{cllc}\hline\hline
\toprule
 Id & $d_{\rm{mod}}$ (pc) & $d_{\rm{par}}$ (pc) & Difference (\%) \\
\midrule
1  & 261 $\pm$ 42  & 264 $\pm$ 35 &  1.2 \\
8  & 142 $\pm$ 23  & 155 $\pm$ 12 &  8.3 \\
13 & 140 $\pm$ 22  & 138 $\pm$  9 &  1.2 \\
16 & 142 $\pm$ 23  & 131 $\pm$  8 &  8.0 \\
17 & 287 $\pm$ 46  & 296 $\pm$ 51 &  3.0 \\
20 & 104 $\pm$ 17  & 152 $\pm$ 12 & 31.3 \\
21 & 158 $\pm$ 25  & 145 $\pm$ 10 &  9.0 \\
31 & 152 $\pm$ 24  & 148 $\pm$ 12 &  2.6 \\
34 & 131 $\pm$ 21  & 147 $\pm$ 10 & 10.8 \\
45 & 738 $\pm$ 118 & ....         &  ... \\
46 & 391 $\pm$ 63  & 283 $\pm$ 47 & 38.0 \\
50 & 143 $\pm$ 23  & 142 $\pm$ 10 &  0.1 \\
4  & 161 $\pm$ 26  &   ....       &  ... \\
5  & 160 $\pm$ 26  &   ....       &  ... \\
10 & 151 $\pm$ 24  &   ....       &  ... \\
14 & 195 $\pm$ 31  &   ....       &  ... \\
35 & 175 $\pm$ 28  & 128 $\pm$ 15 & 36.4 \\
36 & 522 $\pm$ 84  &   ....       &  ... \\
51 & 164 $\pm$ 26  &   ....       &  ... \\
52 & 187 $\pm$ 30  &   ....       &  ... \\
73 & 204 $\pm$ 33  &   ....       &  ... \\
\bottomrule
\end{tabular}
\end{table}


\end{document}